\documentclass[11pt]{article}

\usepackage{amssymb,amsmath,latexsym}
\usepackage{undertilde}
\usepackage[totalwidth=17cm,totalheight=24cm]{geometry}
\newcommand{\C}{{\mathbb C}}

\newcommand{\im}{{\rm i }}
\newcommand{\D}{{\mathcal D }}

\newcommand\be{\begin{eqnarray}}
\newcommand\ee{\end{eqnarray}}

\usepackage{graphicx,ctable,booktabs,verbatim}
\usepackage{float}
\usepackage{setspace}
	\usepackage{multicol}
	
	\usepackage{amsthm} 
\usepackage{setspace}
\theoremstyle{plain}

\theoremstyle{definition}

\theoremstyle{remark}

\usepackage{simplewick}	
	
	\usepackage{mathtools}
\newcommand{\defeq}{\vcentcolon=}

\usepackage{slashed}

\usepackage{tensor}

\begin{document}

\title{Second Order Standard Model}
\author{Johnny Espin and Kirill Krasnov\\
\textit{\small School of Mathematical Sciences}\\
\textit{\small University of Nottingham}}
\date{August 2013}
\maketitle

\begin{abstract} We rewrite the Lagrangian of the fermionic sector of the Standard Model in a novel compact form. The new Lagrangian is second order in derivatives, and is obtained from the usual first order Lagrangian by integrating out all primed (or dotted) 2-component spinors. The Higgs field enters the new Lagrangian non-polynomially, very much like the metric enters the Einstein-Hilbert Lagrangian of General Relativity. We also discuss unification in the second order formalism, and describe a natural in this framework ${\rm SU}(2)\times {\rm SU}(4)$ unified theory.
\end{abstract}

\section{Introduction}

In this paper we carry out an exercise of rewriting the Standard Model (SM) Lagrangian in an unusual way. The SM appears to describe everything in Nature (apart from gravity), but physicists are still far from understanding {\it why} the complicated SM Lagrangian is what it is. The Standard Model is renormalizable, anomaly free, it contains some hints of grand unification, but we are yet to understand other principles underlying its baroque features. One way to make progress in this situation is to collect as many equivalent ways of writing the SM Lagrangian as possible. One of these may then lead to insights that are harder to see in other formulations.

The fermionic sector of the SM is first order in derivatives. The reason for this is partly historical -- every fermion of the SM is described by very much the same Dirac Lagrangian that proved to be so successful for the description of electrons. The more fundamental reason for the omnipresence of the Dirac Lagrangian is that it provides the simplest description of spin $1/2$ particles while exhibiting the manifest hermiticity. 

It is known, however, that fermions can be equivalently described by second order in derivatives Lagrangians. A useful analogy here is between the Hamiltonian and Lagrangian mechanics. The Hamiltonian formulation gives first order evolution equations, but contains twice more independent variables than the second order Lagrangian framework. Similarly, the first order Dirac formalism contains both a fermion field $\psi$, as well as its conjugate momentum (this turns out to be the hermitian conjugate $\psi^\dagger$ of $\psi$). The two fields are treated in the fermionic path integral as independent. As in the passage from Hamiltonian to Lagrangian mechanics one can "integrate out" the momentum variables and arrive at a second order formulation, in a theory of fermions one can start with the first order Dirac Lagrangian and integrate out the "momentum" fields $\psi^\dagger$. One arrives at the second order formulation of fermions. 

The second order formulation of fermions has been studied by many authors. The list of references that we are aware of is \cite{Feynman:1958ty}-\cite{VaqueraAraujo:2012qa}, plus a few more works listed in \cite{Veltman:1997am}. Our approach uses 2-component spinors and is thus closest to that in \cite{Chalmers:1997ui}. As far as we are aware, the second order formalism has only been studied for the case of QED, and the potentially more interesting case of the SM has not been considered. One aim of this paper is to fill this gap. 

Because the second order formulation contains just half of the spinor fields appearing in the Dirac Lagrangian it leads to some simplifications, in particular in the Feynman rules. Thus, as it was pointed out several times in the literature, in the second order formulation the propagator is a multiple of the identity matrix in the spinor space. This should be contrasted with the familiar Dirac propagator, which is a multiple of $(k_\mu \gamma^\mu-m)^{-1}$, and is thus a non-trivial matrix. However, there are also some complications. One of them is that there are new vertices. Thus, even in the simplest case of QED in the second order version, there is a new 4-valent interaction, bilinear in the fermionic field as well as in the electromagnetic potential. The perturbation theory that results from the second order Lagrangian is thus different but produces the same results as the standard one, see e.g. \cite{Veltman:1997am} for a very explicit demonstration. 

Another price to pay for the relative simplicity of the second order formalism is its non-manifest unitarity. Thus, as we shall see below, the procedure of integrating out the fermionic "momentum" fields is most conveniently carried out using the 2-component spinors. Then the Dirac Lagrangian contains 2-component spinors $\psi^A, A=1,2$ as well as their hermitian conjugates $(\psi^\dagger)^{A'}, A'=1,2$. Here $A,A'$ are spinor indices denoting two different types (unprimed, primed) of 2-components spinors. The second order Lagrangian contains only the unprimed spinors $\psi^A$. The manifest hermiticity and thus unitarity of the Dirac Lagrangian is lost. Then, to guarantee unitarity one has to supplement the theory with appropriate reality conditions. The good news is that these reality conditions need only be imposed on the external legs of the diagrams. This has to do with the fact that the second order Lagrangian without any additional reality conditions correctly reproduces all correlation functions of the unprimed fermionic fields. The scattering amplitudes can be extracted from the latter in the usual way, and so the only place where the reality conditions enter is in the determination of the mode decomposition of the free fermionic field. Thus, in the most laborious step of computing the correlation functions the reality conditions play no role whatsoever and can be ignored. The second order Lagrangian itself contains all the information needed for these computations. This is why to a large extent the second order Lagrangian contains all the physics, and the reality conditions are subordinate. Further discussion of hermiticity and unitarity in this formalism can be found in \cite{Veltman:1997am}.

Our main result is an extremely simple second order Lagrangian for the fermionic sector of the SM. It is worth exhibiting the simplicity of this Lagrangian already in the Introduction. It reads
\be\label{L-intr}
\mathcal{L}_{ferm} = -\frac{2}{\rho}D\bar Q^i{D} {Q_i}  -\frac{2}{\rho}D\bar L^i{D} {L_i} - \rho \left( \Lambda \bar Q\right)^i Q_i - \rho \left( \Lambda \bar L\right)^i L_i .
\ee
This should be compared to (\ref{Lferm}). The notations are as follows: 
\be
Q_i=\left(\begin{array}{c} u_i \\ d_i \end{array}\right), \qquad 
L_i=\left(\begin{array}{c} \nu_i \\ e_i \end{array}\right), \qquad 
\left(\Lambda \bar{Q}\right)^i := \left( \Lambda_u^{ij} \bar{u}_j , \Lambda_d^{ij} \bar{d}_j \right), \qquad 
\left(\Lambda \bar{L}\right)^i := \left( \Lambda_\nu^{ij} \bar{\nu}_j , \Lambda_e^{ij} \bar{e}_j \right),
\ee
all fermionic fields $u_i,d_i,\nu_i,e_i,\bar{u}_i,\bar{d}_i,\bar{\nu}_i,\bar{e}_i$ are 2-component unprimed spinors, with $u_i,d_i,\bar{u}_i,\bar{d}_i$ needed to describe quarks, and $\nu_i,e_i,\bar{\nu}_i,\bar{e}_i$ describing leptons. The strong ${\rm SU}(3)$ indices are suppressed, as are the fermionic ones. $\Lambda^{ij}_{u,d,\nu,e}$ are Hermitian $3\times 3$ mass matrices with $i,j=1,2,3$ being the generation index. As we explain in more details below, the doublets $Q_i,L_i$ are ${\rm SU}(2)$ invariant ones constructed by absorbing the Higgs field into the fermions. The covariant derivatives $D$ contain appropriate for each field gauge field, and in the case of $Q_i,L_i$ contain ${\rm SU}(2)$ invariant $W^\pm, Z$ fields constructed from the ${\rm SU}(2)$ gauge field and the Higgs. Note that the right-handed fermions that are uncharged under the weak ${\rm SU}(2)$ got combined into doublets. We have also added the right-handed neutrinos as well as the corresponding mass matrix to account for the neutrino oscillations. This Lagrangian does not contain the Majorana mass terms for the right-handed neutrinos that would be necessary to achieve the seesaw mechanism. We give the corresponding expression in the main text as it is somewhat less elegant. 

The Higgs field $\phi$ enters the second order Lagrangian only via the combination $\rho^2:=|\phi|^2$. The other 3 components of the Higgs doublet have been absorbed into the fields $Q_i,L_i,W^\pm,Z$ to make them ${\rm SU}(2)$ gauge-invariant objects. The idea to use gauge-invariant field combinations is not new, in the context of spontaneous symmetry breaking it has been considered in the literature several times, see \cite{Vlasov:1987vt}-\cite{Masson:2010vx}. We found that the second order formalism naturally leads to this gauge invariant formulation. 

One of its most interesting features of (\ref{L-intr}) is that the Higgs field enters the new Lagrangian non-polynomially, here as a factor of $1/\rho$ in front of the kinetic term. In the main text we shall also give another version of the same Lagrangian where we absorb a factor of $1/\sqrt{\rho}$ into each fermionic field to render the kinetic terms canonical. This keeps the non-polynomical character of the Lagrangian, but now the Higgs enters via a contribution to the covariant derivatives, as well as in the mass terms. 

Due to its non-polynomiality the Lagrangian (\ref{L-intr}) is not manifestly renormalizable, unlike its first order version that is just a $\phi^4$ theory and the renormalizability is evident. The non-polynomiality of the second order Lagrangian in the Higgs field is in a striking resemblence to what one encouters in General Relativity. What is also similar with gravity is the fact that the non-polynomiality  disappears if one passes to the first order formulation. We do not know whether all these similarities are just superficial or there is something deep, but they seem to be worth pointing out. We will further comment on the gravitational analogies in the last section. 

The Lagrangian (\ref{L-intr}), together with the bosonic part that is given in terms of ${\rm SU}(2)$ invariant fields in the main text, can be used as the starting point for computations of the SM scattering amplitudes. As usual this would proceed in two steps. In the first step one would compute correlation functions of all the fields, and in the second the field expansion into modes would be used to extract the scattering amplitudes. It is in this last step where the reality conditions for the free fermionic fields need to be added to the story, but these cause no difficulties. It would be interesting to develop this method of computing the amplitudes further, and possibly even implement it numerically, as it is likely to lead to computational advantages as compared to the rules based on the first order SM Lagrangian. This alternative method of computing the SM amplitudes is another result of the present work. 

Finally, we found it hard to resist to think along the Grand Unification lines, and try to combine the terms in (\ref{L-intr}) into an even more compact expression. Given that the quark and leptons enter the Lagrangian so similarly, it is natural to think of the leptons as the fourth color of quarks, as is done in the Pati-Salam GUT \cite{Pati:1974yy}. Then the 4 terms in (\ref{L-intr}) combine into two, which we can write as
\be\label{L-intr-unif}
\mathcal{L}_{ferm} = -\frac{2}{\rho}D\bar S^i{D} {S_i}  - \rho \bar S_i \Lambda^{ij} S_j,
\ee
where $S_i, \bar{S}^i$ are doublets with an additional index $\alpha=1,2,3,4$ whose first 3 values enumerate the color and $\alpha=4$ corresponds to leptons. The would be ${\rm SU}(4)$ symmetry is broken by the mass matrix (with a complicated structure, see below) in the last term. It is notable that in this second order formulation all fermions of the SM got combined into an 8-dimensional 2-component spinor multiplet $S_i$, plus a multiplet that transforms under a complex conjugate representation of the unbroken gauge group ${\rm U}(1)\times {\rm SU}(3)$. We give further details on this Pati-Salam-like Lagrangian in the main text. 

The organization of this paper is as follows. We start in Sec. \ref{sec:QED} by a quick reminder of how the usual Weyl, Majorana and Dirac actions look like in 2-component spinor notations. This fixes our conventions and notations. In this section we also carry the exercise of integrating out the primed spinor fields for the Dirac Lagrangian, thus arriving at the Lagrangina of second order QED. Section \ref{sec:SM} reviews the fermionic sector of the SM in terms of 2-component spinors. Section \ref{sec:SOL} is the main one where we obtain the second order Lagrangian (\ref{L-intr}). We also analyze the case with the Majorana mass term here. In Sec. \ref{sec:bosonic} we write the bosonic sector Lagrangian in terms of the same gauge-invariant variables that we were led to introduce in the fermionic sector. The material here has appeared in the literature before, and we give it for completeness. Section \ref{sec:inter} describes some simplest interaction vertices that follow from our Lagrangian, to get a better feel of how the second order perturbation theory works. We discuss unification in Section \ref{sec:unific} and finish with a discussion. 

\section{Preliminaries: Second order QED}
\label{sec:QED}

We start by considering a much simpler setup of Quantum Electrodynamics (QED). To establish our conventions, and to prepare for integration over the primed spinors, let us write the familiar fermionic Lagrangians in terms of 2-component spinors. We follow the 2-component spinor conventions of \cite{TorresGomez:2012sr}.

\subsection{Single massless Weyl fermion}

The Lagrangian for a single massless Weyl fermion reads:
\be\label{weyl-massless}
{\cal L}_{\text{Weyl}}= -\im \sqrt{2} (\lambda^\dagger)_{A'} \theta^{\mu\,A'A} \partial_\mu \lambda_A\equiv -\im \sqrt{2} \lambda^\dagger \partial \lambda.
\ee
Here $\lambda_A$ is a 2-component spinor, $\lambda_A^{\dagger}$ is its Hermitian conjugate  and $\theta_\mu^{A'A}$ is the soldering form
\be
\theta^A_{\mu A'} \theta_{\nu A}{}^{A'} = \eta_{\mu\nu},
\ee
where $\eta_{\mu\nu}={\rm diag}(-1,1,1,1)$. We have also introduced the notation $\partial^{AA'} := \theta^\mu \partial_\mu^{AA'}$ and written the Lagrangian in an index-free way. Our index-free convention is that the unprimed fermions are always contracted as $\lambda\chi\equiv \lambda^A \chi_A$, while the primed fermions are contracted in the opposite way $\lambda^\dagger \chi^\dagger\equiv (\lambda^\dagger)_{A'}(\chi^\dagger)^{A'}$. The factor of $\sqrt{2}$ is introduced for convenience, and the minus in front of the kinetic term is convention dependent. With our conventions it is needed to get the positive-definite Hamiltonian. The Lagrangian is Hermitian, modulo a surface term. 

\subsection{Majorana mass term}

When fermions are Grassmann valued we can have the Majorana mass term. Thus, consider
\be\label{majorana}
{\cal L}_{\text{Majorana}}=-\im \sqrt{2} \lambda^\dagger \partial \lambda - (m/2)\lambda\lambda - (m/2)\lambda^\dagger\lambda^\dagger,
\ee
where we have used the index-free notation, and $m$ is the parameter with dimensions of mass, later to be identified with the physical mass. Note that we need to add both terms in order for the Lagrangian to be Hermitian.

\subsection{Dirac fermions}

Dirac fermions are obtained by taking two massive Weyl fermions of equal mass. The system is then invariant under ${\rm SO}(2)$ rotations mixing the fermions. Since ${\rm SO}(2)\sim {\rm U}(1)$, complex linear combinations of fermions can be introduced and the Lagrangian rewritten in an explicitly ${\rm U}(1)$-invariant way. This symmetry can be made local by introducing a ${\rm U}(1)$ gauge field and converting the usual derivative to the covariant one. Thus, we define
\be\label{cov-ders}
D_\mu \xi = (\partial_\mu - \im e A_\mu)\xi, \qquad
D_\mu \chi = (\partial_\mu + \im e A_\mu)\chi,
\ee
where $A_\mu$ is the electromagnetic potential and $e$ is the electron charge. Note that, since the fields $\xi,\chi$ are charged in the opposite way, the expressions for the covariant derivatives on these fields differ by a sign in front of $A_\mu$. The gauge transformation rules are: for the fermions $\xi\to e^{\im\phi} \xi, \chi\to e^{-\im\phi}\chi$, and for the electromagnetic potential $A_\mu\to A_\mu-(1/e) \partial_\mu \varphi$. The Lagrangian becomes
\be\label{dirac-a}
{\cal L}_{\text{Dirac}}=- \im \sqrt{2} \xi^\dagger D \xi-\im \sqrt{2} \chi^\dagger D \chi - m\chi\xi -m\xi^\dagger \chi^\dagger,
\ee
where as before $D:=\theta^\mu D_\mu$. 

\subsection{Chiral Dirac theory}

At the level of the path integral we can integrate out the fermionic fields $\xi^\dagger, \chi^\dagger$ and obtain a chiral Lagrangian involving only unprimed spinors. From field equations for the primed spinors we get:
\be\label{eqs-dirac}
\xi^\dagger = -\frac{\im\sqrt{2}}{m} D \chi, \qquad \chi^\dagger = -\frac{\im\sqrt{2}}{m} D \xi.
\ee
Substituting this into the Lagrangian (\ref{dirac-a}) we get:
\be\label{L-D-chiral}
{\cal L}_{\text{chiral}} = - \frac{2}{m} D \chi  D \xi  - m \chi \xi.
\ee
As the final step, the kinetic term here can be put into the canonical form by doing the rescaling
\be
\xi\to \sqrt{m}\xi, \qquad \chi\to \sqrt{m} \chi.
\ee
The new spinor fields have mass dimension one, and the Lagrangian takes the following simple form
\be
{\cal L}_{\text{chiral}} = - 2 D \chi  D \xi  - m^2 \chi \xi \equiv - 2 D_{A'}{}^{A} \chi_A  D^{A' B} \xi_B  - m^2 \chi^A \xi_A,
\ee
where we reintroduced the spinor indices to make everything explicit. Our aim is to write the SM fermionic sector in a similar fashion.

\section{Preliminaries: SM Lagrangian via 2-component spinors}
\label{sec:SM}

We follow \cite{Dreiner:2008tw}, with some differences in conventions. 

\subsection{Standard Model fermions}

The SM fermions can be put together in the following table

\bigskip
\begin{tabular}{c c c c c c}
2-component fermions & SU(3) & SU(2) & $Y$ & $T_3$ & $Q=T_3+Y$ \\
\hline
$Q_i = \left(\begin{array}{c} u_i \\ d_i \end{array} \right)$ & $\begin{array}{c} {\rm triplet}  \\ {\rm triplet} \end{array}$ & doublet & $\begin{array}{c} 1/6  \\ 1/6\end{array}$ & $\begin{array}{c} 1/2  \\ -1/2\end{array}$ & $\begin{array}{c} 2/3  \\ -1/3\end{array}$ \\
$\bar{u}_i$ & triplet & singlet & $-2/3$ & 0 & $-2/3$ \\
$\bar{d}_i$ & triplet & singlet & $1/3$ & 0 & $1/3$ \\
\hline
$L_i = \left(\begin{array}{c} \nu_i \\ e_i \end{array} \right)$ & $\begin{array}{c} {\rm singlet}  \\ {\rm singlet} \end{array}$ & doublet & $\begin{array}{c} -1/2  \\ -1/2\end{array}$ & $\begin{array}{c} 1/2  \\ -1/2\end{array}$ & $\begin{array}{c} 0  \\ -1\end{array}$ \\
$\bar{e}_i$ & singlet & singlet & $1$ & 0 & $1$ \\
$\bar{\nu}_i$ & singlet & singlet & $0$ & 0 & $0$ \\
\hline
\end{tabular}

\bigskip
All fermionic fields here are unprimed 2-component spinors. Thus, $\bar{u}_i$ is {\it not} the Hermitian conjugate of $u_i$, with the latter being denoted by $u_i^\dagger$. The first 4 fermionic fields are the ones needed to describe quarks, the last 4 are those for describing leptons. Note that the last line is not a part of the Standard Model, but is now believed to be needed to account for neutrino masses. Also note that the strong ${\rm SU}(3)$ triplet fields are actually a collection of 3 different two-component spinors. Thus, e.g. $u_i$ has two types of indices suppressed: the usual spinor index, as well as the strong ${\rm SU}(3)$ index. With no index suppression this field would be denoted by $u_{i A}^\alpha$, where $A=1,2$ is the usual spinor index, and $\alpha=1,2,3$ is the index on which ${\rm SU}(3)$ acts. The only index that is indicated explicitly is that enumerating the generations $i=1,2,3$. It is an instructive exercise to count the number of 2-component fermionic fields in the above table. We have $4\times 3=12$ for the quark sector and $4$ for the lepton sector. Thus, we need 16 2-component spinors for each generation of the Standard Model.

\subsection{Higgs field}

This is the field that plays the central role in the the Standard Model. It is a complex field of ${\rm U}(1)$ charge $Y=1/2$. It is also a weak ${\rm SU}(2)$ doublet, i.e. it can be written as a column

\bigskip
\begin{tabular}{c c c c c c}
Higgs & SU(3) & SU(2) & $Y$ & $T_3$ & $Q=T_3+Y$ \\
\hline
$\phi = \left(\begin{array}{c} \phi^+ \\ \phi^0 \end{array} \right)$ & singlet & doublet & $\begin{array}{c} 1/2  \\ 1/2\end{array}$ & $\begin{array}{c} 1/2  \\ -1/2\end{array}$ & $\begin{array}{c} 1  \\ 0\end{array}$ \\
\end{tabular}

\bigskip
Note that being an ${\rm SU}(2)$ doublet, it is really a collection of 2 complex fields $\phi^+$ (the plus denotes the fact that the electric charge $Q=+1$) and $\phi^0$ (with $Q=0$). Alternatively, it is a collection of 4 real fields. We shall denote the weak ${\rm SU}(2)$ index by $a, b, \ldots = 1,2$. Thus we can write the Higgs field as $\phi_a$, with $\phi_1=\phi^+$ and $\phi_2=\phi^0$.

\subsection{Fermionic sector of the Standard Model}

Using index-free notations, the Lagrangian for the fermionic sector of the Standard Model reads:
\begin{align}\begin{split}
\mathcal{L}_{ferm} = ~& -i\sqrt{2}Q^{\dagger i}{D} {Q_i} - i\sqrt{2}\bar u^{\dagger i}{D} \bar u_i- i\sqrt{2}\bar d^{\dagger i}{D} \bar d_i- i\sqrt{2}L^{\dagger i}{D} {L_i}-i\sqrt{2} \bar e^{\dagger i}{D} {\bar e_i}- i\sqrt{2}\bar \nu^{\dagger i}{D} \bar \nu_i \\
&+ Y_u^{ij}  \phi^T \varepsilon Q_i \bar u_j - Y_d^{ij} \phi^\dagger Q_i \bar d_j +Y_\nu^{ij} \phi^T \varepsilon L_i \bar \nu_j - Y_e^{ij} \phi^\dagger L_i \bar e_j \\
&- (Y_u^\dagger)^{ij} \bar u_i^\dagger Q^\dagger_j\varepsilon  \phi^*   - (Y_d^\dagger)^{ij}\bar d_i ^\dagger Q_j  ^\dagger\phi -(Y_\nu^\dagger)^{ij} \bar \nu_i^\dagger L^\dagger_j\varepsilon  \phi^*   - (Y_e^\dagger)^{ij}\bar e_i ^\dagger L_j  ^\dagger\phi \\
& -\frac{1}{2}M^{ij}_{\bar\nu} \bar\nu_i\bar\nu_j - \frac{1}{2}(M_{\bar\nu}^\dagger)^{ij} \bar\nu_i^\dagger\bar\nu_j^\dagger .
\end{split}\label{Lferm}\end{align}
Here as before $D^{AA'} \equiv \theta^{\mu AA'}D_\mu$, where $D_\mu$ is an appropriate for each field covariant derivative. The quantities $Y^{ij}$ are arbitrary complex $3\times 3$ mass matrices. All fermions are Grassmann-valued. 

We have given a formulation of the SM with the right-handed neutrinos added in. To get the usual SM one just has to erase all the terms containing the $\bar{\nu}$ field. Equivalently, it is obtained by setting to zero the mass matrices $Y_\nu^{ij}, M_\nu^{ij}$. 

Only the ${\rm SU}(2)$ index structure in the above Lagrangian needs clarifications, all other suppressed indices are self-explanatory. Our ${\rm SU}(2)$ index conventions are as follows. As we have already mentioned, the Higgs field is considered to be an object $\phi_a$ with a single ${\rm SU}(2)$ index in the lower position. Its transpose is then an object $(\phi^T)^a$. The complex conjugate field $(\phi^*)_a$ still carries a lower position index, while the Hermitian conjugate is $(\phi^\dagger)^a$. Similarly, the quark doublet $Q_a$ has a lower index position. Its Hermitian conjugate is an object $(Q^\dagger)^a$. The quantity $\epsilon\equiv \epsilon_a{}^b$ is the matrix 
\be
\epsilon_a{}^b=\left(\begin{array}{cc} 0 & 1 \\-1 & 0 \end{array} \right).
\ee
Then the object $\phi^T \epsilon Q \equiv (\phi^T)^a \epsilon_a{}^b Q_a$ is invariant under the action of ${\rm SL}(2,\C)$ via $Q\to gQ, \phi\to g\phi$ since $g^T \epsilon g=\epsilon$. In particular, $\phi^T \epsilon Q$ is ${\rm SU}(2)$ invariant. It is then clear that all the terms in the second line of (\ref{Lferm}) are ${\rm SU}(2)$ invariant. The ${\rm U}(1)$ invariance is also easily checked by noting the $Y$-charges of the objects. The third line in (\ref{Lferm}) is then the Hermitian conjugate of the second. The last line is the Majorana mass terms for the right-handed neutrinos. It is also worth noting that the position of the generation indices (upper or lower) does not matter, as all these indices are explicitly indicated. We put them wherever convenient so as not to clatter the formulas. 

\section{Second order formulation of the Standard Model}
\label{sec:SOL}

We now carry out the procedure of integrating out the primed 2-component spinors from the significantly more involved SM Lagrangian. 

\subsection{Quark sector}
We start with the quark sector as there is no Majorana mass term in this case. The equations of motion for the unprimed spinors are:

\begin{spacing}{2}
\begin{align}\begin{split}
\begin{array}{|c|cl|}
\hline  Q_i^{\dagger}: & i\sqrt{2}D Q^i =& -\left( \epsilon\phi^*\right) \bar{u}^\dagger_j (Y_u^\dagger)^{ji}- \phi ~\bar{d}^\dagger_j (Y_d^\dagger)^{ji} \\  
\bar u_i^{\dagger}: & i\sqrt{2}D\bar u^i =& - (Y_u^\dagger)^{ij}Q^\dagger_j\left( \epsilon\phi^*\right) \\
\bar{d} _i^\dagger: &i\sqrt{2} D\bar d^i =&- (Y_d^\dagger)^{ij}Q^\dagger_j~\phi \\ \hline
\end{array}\end{split}
\end{align}
\end{spacing}

We now notice that some structure is making itself explicit in the equations of motion. Thus, let us combine the components of the Higgs field into the following $2\times 2$ matrix:
\begin{align}
\rho\Phi^\dagger \defeq\left( \epsilon\phi^*, \phi\right ) \equiv  \left( \begin{array}{cc}
(\phi^0)^* & \phi^+ \\ - \phi^- & \phi^0\\
\end{array}\right).
\end{align}
Under the weak $SU(2)$ the matrix $\Phi^\dagger$ transforms as:
\begin{align}
\Phi^\dagger  ~\mapsto~ \Omega\Phi^\dagger,
\end{align}
while the field $\rho$ remains invariant. It is clear that $\rho^2$ is just the modulus squared of the Higgs field. 

To make the above equations look more transparent, we define the new quark singlets as linear combinations of the old ones:
\begin{align}\label{new-bars}
\bar{u}^i ~\rightarrow~ (Y_u^\dagger)^{ij}\bar{u}_j, \quad \bar{d}^i ~\rightarrow~ (Y_d^\dagger)^{ij}\bar{d}_j.
\end{align}
It is clear that this is designed to eliminate the mass matrices from the last pair of equations. Note that this is a constant reparameterization of the fields, so it does not lead to any change in the derivative operators. We further combine the new quark singlets into a row
\be
\bar{Q} _i &\defeq \left( \bar{u}_i ~,~ \bar{d}_i\right)
\ee
In terms of the new quark singlets the equations of motion become:
\begin{spacing}{2}
\begin{align}\begin{split}
\begin{array}{|c|cl|}
\hline  Q _i^\dagger: & i\sqrt{2}D Q_i =& -\rho~\Phi^\dagger \left( \bar{Q}^\dagger\Lambda\right)_i\\  
\bar{Q} _i^\dagger: &i\sqrt{2} D\bar Q_i =&- \rho~ Q^\dagger_i \Phi^\dagger\\ \hline
\end{array}\end{split}
\end{align}
\end{spacing}
\noindent Here we introduced new hermitian mass matrices
\be
\Lambda_q^{ij} &\defeq Y_q^{ik}(Y_q^\dagger)^{kj},
\ee
as well as a new column
\be\label{Q-Lambda}
\left(\bar{Q}^\dagger \Lambda\right)^i \equiv  \left(\begin{array}{c} \bar{u}^\dagger_j \Lambda_u^{ji} \\ \bar{d}^\dagger_j \Lambda_d^{ji}
\end{array}   \right)
\ee

While introducing the new doublet $\bar{Q}_i$ has made the equations look more symmetric, there is no complete symmetry. Indeed, the doublet $Q_i$ transforms under the weak ${\rm SU}(2)$, and so does the Higgs matrix $\Phi^\dagger$, while $\bar{Q}_i$ does not transform. This suggests that we define a new set of ${\rm SU}(2)$-invariant quark variables $\Phi Q_i$
\begin{align}\label{new-Q}
\Phi Q_i \defeq Q_i^{inv}.
\end{align}
This is a Higgs-field dependent ${\rm SU}(2)$ gauge rotation of the original quark doublet. As such, it can be pulled through the derivative as long as one takes into account that the gauge field also transforms. As we will work out in details below, the new gauge field will be an ${\rm SU}(2)$-invariant object. Keeping in mind this change in the derivative operator we can write the field equations as:
\begin{spacing}{2}
\begin{align}\begin{split}
\begin{array}{|c|cl|}
\hline  Q _i^\dagger: & i\sqrt{2}D Q_i =& -\rho~\left(\bar{Q}^\dagger \Lambda\right)_i \\  
\bar{Q} _i^\dagger: &i\sqrt{2} D\bar Q_i =&- \rho~ Q^{\dagger}_i \\ \hline
\end{array}\end{split}\label{realquarks}
\end{align}
\end{spacing}
\noindent We have dropped the superscript $inv$ from the $Q_i$ to avoid clattering of the notations. We see that the equations become much simpler than in terms of the original fields. 

We now substitute the primed spinors obtained from the above field equations into the Lagrangian (\ref{Lferm}) and obtain the following second-order Lagrangian:
\begin{align}\label{L2-quarks}
\mathcal{L}_{quarks}= -\frac{2}{\rho}D\bar Q^i{D} {Q_i} - \rho \left( \Lambda \bar Q\right)^i Q_i ,
\end{align}
where we have introduced a new row
\be
\left(\Lambda \bar{Q}\right)^i := \left( \Lambda_u^{ij} \bar{u}_j , \Lambda_d^{ij} \bar{d}_j \right),
\ee
which is the Hermitian conjugate of (\ref{Q-Lambda}) $\left(\Lambda \bar{Q}\right)^\dagger=\bar{Q}^\dagger \Lambda$. The calculation that leads to (\ref{L2-quarks}) simplifies significantly by noting that half of the kinetic terms cancels the mass terms for the primed spinors, while the other half survives. Then the kinetic term in (\ref{L2-quarks}) is easily obtained from  the kinetic term $Q^{\dagger i} D Q_i$ by substituting the expression for $Q^{\dagger i}$. The mass term in (\ref{L2-quarks}) is easily obtained by combining the mass terms for the unprimed spinors in (\ref{Lferm}), and taking into account the definitions (\ref{new-bars}), (\ref{new-Q}) of the new fermionic variables. The derivative operator acting on $Q_i$ in (\ref{L2-quarks}) takes into account the field redefinition (\ref{new-Q}). 

The Lagrangian (\ref{L2-quarks}) is much simpler than the terms in (\ref{Lferm}) from which it was obtained, containing just half the terms from the quark sector. However, it is clearly non-polynomical in the Higgs scalar field $\rho$, because of the presence of $1/\rho$ in the kinetic term. This is exactly the same phenomenon as one sees in (\ref{L-D-chiral}). In the case of Dirac theory we have rescaled the fermionic fields to give them mass dimension one, and brought the kinetic term into the canonical form. This suggests that we should do the same in (\ref{L2-quarks}). However, $\rho$ is now a dynamical field. Absorbing it into the fermion fields thus changes the derivative operators acting on both $\bar{Q}_i, Q_i$. Denoting the new Higgs-containing derivative operators by the curly ${\mathcal D}$ we finally write:
\begin{align}\label{L2-quarks*}
\mathcal{L}^{(2)}_{quarks} =  -2\mathcal{D}\bar Q^i\mathcal{D} {Q_i} - \rho^2 \left(\Lambda \bar Q\right)^i Q_i 
\end{align}
where $1/\sqrt{\rho}$ was absorbed into each spinor field. The new covariant derivative $\mathcal{D}$ contains non-polynomial Higgs-quarks interactions as well as the physical $SU(2)$-frozen gauge fields when acting on the unbarred doublet. Expanding (\ref{L2-quarks*}) around the Higgs VEV $\rho\to v+\rho$ one gets the free massive quarks with masses being multiples of the eigenvalues of the hermitian mass matrices $\Lambda_q^{ij}$, together with quark interactions with the gauge fields as well as the Higgs. We will spell out the simplest interactions below. It is clear that interaction vertices with the Higgs can be of arbitrarily high valency (due to non-polynomiality in $\rho$). 

The field equations (\ref{realquarks}) for the new fermionic fields of mass dimension one read
\be\label{quarks-RC}
i\sqrt{2}{\mathcal D} Q_i = -\rho~\left(\bar{Q}^\dagger \Lambda\right)_i , \qquad i\sqrt{2} {\mathcal D}\bar Q_i =- \rho~ Q^{\dagger}_i.
\ee
These are now to be interpreted as the reality conditions, whose linearised versions are to be imposed on the external lines. 

\subsection{Leptonic sector without the Majorana mass terms}

Let us first set all the Majorana mass terms to zero. In this case the analysis is exactly the same as in the previous subsection. Thus, introducing the new barred fermion fields 
\begin{align}\label{new-bars-l}
\bar{\nu}^i ~\rightarrow~ (Y_\nu^\dagger)^{ij}\bar{\nu}_j, \quad \bar{e}^i ~\rightarrow~ (Y_e^\dagger)^{ij}\bar{e}_j.
\end{align}
we put the new fields together in a row
\be
\bar{L} _i \defeq \left( \bar{\nu}_i ~,~ \bar{e}_i\right),
\ee
and define another row 
\be
\left(\Lambda \bar{L}\right)^i := \left( \Lambda_\nu^{ij} \bar{\nu}_j , \Lambda_e^{ij} \bar{e}_j \right),
\ee
where $\Lambda_l = Y_l Y_l^\dagger$ are the Hermitian mass matrices. We also define the physical ${\rm SU}(2)$-invariant unbarred leptonic doublet $L^{inv}_i = \Phi L_i$. Writing everything in terms of these quantities we get the following Lagrangian
\begin{align}\begin{split}
\mathcal{L}_{leptons} =  -i\sqrt{2}L^{\dagger i}{D} {L_i} - i\sqrt{2}\left( D \bar  L^i \right) \left(\bar L^\dagger \Lambda\right)_i - \rho \left( \Lambda \bar L\right)^i L_i -  \rho \, L^{\dagger i} \left( \bar L^\dagger \Lambda\right)_i .
\end{split}\end{align}
The resulting equations for the primed spinors are 
\begin{spacing}{2}
\begin{align}\begin{split}
\begin{array}{|c|cl|}
\hline  L _i^\dagger: & i\sqrt{2}D L_i =& -\rho~\left(\bar{L}^\dagger \Lambda\right)_{i}\\  
\bar{\nu} _i^\dagger: &i\sqrt{2} D\bar L_i =&- \rho~ L^{\dagger}_{i}  \\\hline
\end{array}\end{split}\label{feqs-leptons}
\end{align}
\end{spacing}
\noindent Substituting the resulting primed spinors into the Lagrangian we get
\begin{align}\begin{split}
\mathcal{L}_{leptons}=  -\frac{2}{\rho} D \bar L^i D {L_i} - \rho \left(\Lambda \bar L\right)^i L_i,
\end{split}\label{L2-leptons}\end{align}
One can now rescale the lepton fields to give them mass dimension one and convert the kinetic terms into a standard form. One obtains a Lagrangian as in (\ref{L2-quarks*}). The sum of (\ref{L2-leptons}) and (\ref{L2-quarks}) is the Lagrangian quoted in the Introduction. 

\subsection{Majorana mass terms}

We now reinstate the Majorana mass terms. This leads to a more complicated analysis and not so simple final result. Performing the same redefinitions of the fermionic variables as was done above, we can write the original Lagrangian in terms of the new spinor fields:
\begin{align}\begin{split}
\mathcal{L}_{leptons} = ~& -i\sqrt{2}L^{\dagger i}{D} {L_i} - i\sqrt{2}\left( D \bar  L^i \right) \left(\bar L^\dagger \Lambda\right)_i - \rho \left( \Lambda \bar L\right)^i L_i -  \rho \, L^{\dagger i} \left( \bar L^\dagger \Lambda\right)_i \\
&-\frac{1}{2}(Y_\nu^\dagger)^{ik}(Y_\nu^\dagger)^{jl}M^{ij}_{\bar\nu}\bar\nu_k\bar\nu_l - \frac{1}{2}Y_\nu^{ki}Y_\nu^{lj}(M_{\bar\nu}^\dagger)^{ij}\bar\nu_k^\dagger\bar\nu^\dagger_l .
\end{split}\end{align}
The structure of the last two terms suggests the following redefinition of the right-handed neutrino mass matrix
\begin{align}
 (Y_\nu^\dagger)^{ik}(Y_\nu^\dagger)^{jl}M^{ij}_{\bar\nu}~\rightarrow~M^{kl}_{\bar\nu}.
\end{align}
The new mass matrix is still symmetric. The Lagrangian becomes
\begin{align}\begin{split}
\mathcal{L}_{leptons} =  &-i\sqrt{2}L^{\dagger i}{D} {L_i} - i\sqrt{2}\left( D \bar  L^i \right) \left(\bar L^\dagger \Lambda\right)_i - \rho \left( \Lambda \bar L\right)^i L_i -  \rho \,L^{\dagger i}\left( \bar L^\dagger \Lambda\right)_i  \\
&-\frac{1}{2}M^{ij}_{\bar\nu}\bar\nu_i\bar\nu_j - \frac{1}{2}(M_{\bar\nu}^\dagger)^{ij}\bar\nu_i^\dagger\bar\nu^\dagger_j .
\end{split}\label{Lleptons}\end{align}
The resulting equations of motion for the primed spinors are as follows
\begin{spacing}{2}
\begin{align}\begin{split}
\begin{array}{|c|cl|}
\hline  L _i^\dagger: & i\sqrt{2}D L_i =& -\rho~\left(\bar{L}^\dagger \Lambda\right)_{i}\\  
\bar{\nu} _i^\dagger: &i\sqrt{2} D\bar \nu^i =&- \rho~ \nu^{\dagger i} - \bar\nu^{\dagger j} (M_{\bar\nu}^\dagger)_{jk}(\Lambda^{-1})^{ki}\\ 
\bar{e} _i^\dagger: &i\sqrt{2} D\bar e_i =&- \rho~ e^\dagger_i  \\\hline
\end{array}\end{split}\label{realleptons2}
\end{align}
\end{spacing}
We can now solve for the barred primed spinors using the first equation. From this we in particular get $\bar{\nu}^\dagger$, which can be substituted into the second equation. The last pair is then solved for $L^\dagger$ fermions. 

After the solutions are substituted into the Lagrangian, simplifications result. Thus, it is easy to note that the first and the last terms in the first line of (\ref{Lleptons}) cancel each other in view of the first equation in (\ref{realleptons2}). Indeed, we can combine these two terms as
\begin{align}
\mathcal{L}_{leptons} \supset L^\dagger_i\left(-i\sqrt{2}{D} {L_i} -  \rho  \left(\bar L^\dagger\Lambda\right)_i \right).
\end{align}
It is clear that this combination gives zero on the first equation in (\ref{realleptons2}). To eliminate the remaining primed spinors we again need just the first field equation that gives us $\bar{L}^\dagger$. Overall, this gives:
\begin{align}\begin{split}
\mathcal{L}_{leptons}^{(2)}= ~& -\frac{2}{\rho} D \bar L^i D {L_i} - \rho \left(\Lambda \bar L\right)^i L_i \\
&-\frac{1}{2}M_{\bar\nu}^{ij}\bar\nu_i\bar\nu_j + \frac{1}{\rho^2}(\Lambda_\nu^{-1}M_{\bar\nu}^\dagger\Lambda_\nu^{-1})^{ij} (D L_i)^\nu (D L_j)^\nu,
\end{split}\label{Llep2nd}\end{align}
where $(DL_i)^\nu$ stands for the first $\nu$-component of the doublet $DL_i$. 

This is a complicated Lagrangian. To understand what is going on here it is useful to return to the second field equation in (\ref{realleptons2}). Substituting the solution for $\bar{\nu}^\dagger_i$ from the first equation one gets:
\be
\im\sqrt{2}\left( D\bar{\nu}^i - \frac{1}{\rho} (\Lambda_\nu^{-1}M_{\bar\nu}^\dagger\Lambda_\nu^{-1})^{ij} (D L_j)^\nu\right) = -\rho \, \nu^{\dagger i}.
\ee
If one expands all terms in this equation around the Higgs VEV $\langle \rho\rangle = v$, the terms linear in the fields are
\be
\im\sqrt{2} \, \partial \left( \bar{\nu}^i - \frac{1}{v} (\Lambda_\nu^{-1}M_{\bar\nu}^\dagger\Lambda_\nu^{-1})^{ij} \nu_j \right) = -v \, \nu^{\dagger i}.
\ee
This equation suggests that we should introduce a new barred neutrino field
\be\label{nubar-new}
\bar{\nu}^{new}_i := \bar{\nu}_i - \frac{1}{v} (\Lambda_\nu^{-1}M_{\bar\nu}^\dagger\Lambda_\nu^{-1})_{ij} \nu^j,
\ee
as it is this field that satisfies reality conditions similar to those for all other 2-component fermions present. One should then rewrite the linearisation of the Lagrangian (\ref{Llep2nd}), as well as the interaction vertices in terms of $\bar{\nu}^{new}_i$, and proceed with the usual perturbation theory calculations. It is clear that the result is complicated, as it relies on the field redefinition (\ref{nubar-new}) that in turn relies on the Higgs assuming its VEV. Thus, it seems that no elegant second order Lagrangian is possible when the Majorana mass terms are present. Of course, one can still work with the Lagrangian that is not elegant, but then one may as well work with the usual first order Lagrangian (\ref{Lferm}) where the Majorana mass terms are not more unnatural than any other terms present. 

Thus, if just for a moment one takes the viewpoint that the second order Lagrangian is more fundamental than the usual first order one, one seems to be led to the conclusion that the simple Lagrangian (\ref{L-intr}) is all one has. This Lagrangian explains neutrino oscillations by giving the neutrinos masses, but does not by itself explain why these masses are so small as compared to the other ones. Without the Majorana mass terms there is also not enough room to explain other SM mysteries. Indeed, with the Majorana mass terms the lepton number is no longer conserved, and this plays an important role in many beyond the SM scenarios. In particular, the Majorana mass terms are of importance in the $\nu$MSM explanation \cite{Asaka:2005pn} of the baryon asymmetry of the Universe.

One way out of this could be to add to the Lagrangian (\ref{L-intr}) terms that are natural from the second order point of view, but spoil the renormalizability of the theory. This can provide new couplings which can then in turn be useful for the beyond the Standard Model scenarios. We will not attempt any of this in the present paper, and continue to explore the second order Lagrangian without the Majorana terms. 

\section{Bosonic sector in terms of gauge-invariant variables}
\label{sec:bosonic}

In this section we rewrite the bosonic sector of the Standard Model in terms of the same gauge-invariant fields that were used in our fermionic Lagrangian. The idea that this is possible and, in fact, gives a much more clear perspective on the "symmetry breaking" mechanism is not new, as far as we know was first used in \cite{Vlasov:1987vt} in the context of the Hamiltonian formulation and more recently in \cite{Chernodub:2008rz,Faddeev:2008qc,Masson:2010vx}. Our analysis here basically repeats \cite{Faddeev:2008qc} with minor differences in conventions. We give it here for completeness, as well as to be able to spell out some simple interaction vertices that follow from our formalism in the next section.

\subsection{Higgs sector}

We denote the gauge fields associated to the Standard Model group $SU(2)\times U(1)$\footnote{We omit the $SU(3)$ part as it does not affect the discussion that follows.} by $B_\mu$ and $Y_\mu$ respectively and their coupling constants by $g_2, ~g_1$.

We first repeat the construction of an ${\rm SU}(2)$ element out of the Higgs field. This is the same construction that already appeared in the previous sections. We start with the Higgs field that is an $SU(2)$ doublet with the $Y$-charge of $1/2$. The covariant derivative of the Higgs reads
\be
D_\mu \phi = \partial_\mu \phi + \im g_2 B_\mu \phi + \frac{\im g_1}{2} Y_\mu \phi,
\ee
where $B_\mu = T^a B_\mu^a$ and $T^a=(1/2)\sigma^a$, where $\sigma^a$ are the usual Pauli matrices. The ${\rm SU}(2)$ connection $B_\mu$ transforms as $B_\mu\to \Omega^\dagger B_\mu \Omega + (1/\im g_2) \Omega^\dagger (\partial_\mu \Omega)$, and the ${\rm U}(1)$ connection as $Y_\mu\to Y_\mu + (1/g_1) \partial_\mu \xi$. 

We now parametrize this doublet as
\begin{align}
\phi \equiv \rho \chi, \qquad \rho \in \mathbb{R}^+, ~\chi \in \mathbb{C}^2 ~{\rm with}~|\chi|^2 =1.
\end{align}
Using the spinor metric $\varepsilon$, we can construct
\begin{align}
\Phi \equiv \left(\begin{array}{c} (\varepsilon \chi)^T \\ \chi^\dagger \end{array} \right)\in SU(2)
\end{align}
This ${\rm SU}(2)$-valued object transforms under the Standard Model gauge group as follows:
\begin{align}
U(1)~&:~ \Phi ~\mapsto ~ e^{i\xi T^3}\Phi  \\
SU(2)~&:~ \Phi ~\mapsto ~ \Phi\Omega^\dagger
\end{align}
We can then define a covariant derivative operator such that $D_\mu \Phi$ transforms covariantly. This derivative operator is given by:
\begin{align}\label{d-Phi}
D_\mu \Phi := \partial_\mu \Phi - i g_2 \Phi B_\mu  +i g_1Y_\mu T^3 \Phi =  \left(\begin{array}{c} (\varepsilon D_\mu \chi)^T \\ (D_\mu \chi)^\dagger \end{array} \right),
\end{align}
where to obtain the last expression we have used $\epsilon B_\mu = B_\mu^T \epsilon^T$ which can be checked to hold for all 3 generators $T^a$. 

We can now use the object $\Phi$, as well as its covariant derivative (\ref{d-Phi}) to rewrite the Higgs kinetic term $|D_\mu \phi|^2$ as
\begin{align}
|D_\mu \phi |^2 = \left( \partial_\mu \rho \right)^2 + \rho^2 |D_\mu \chi |^2 = \left( \partial_\mu \rho \right)^2 + \frac{\rho^2}{2}{\rm Tr}  |D_\mu \Phi|^2,
\end{align}
where we have defined
\begin{align}
|\phi|^2 = \rho^2.
\end{align}

Now let us recall the construction of ${\rm SU}(2)$-invariant doublets. In (\ref{new-Q}) we have defined $Q_i^{new}$ so that $Q_i = \Phi^\dagger Q_i^{new}$. We then rewrote the Lagrangian in terms of $Q_i^{new}$ and the gauge-transformed ${\rm SU}(2)$ connection
\begin{align}
W_\mu \defeq \Phi B_\mu \Phi^\dagger + \frac{1}{ig_2}\Phi \partial_\mu \Phi^\dagger = 
\Phi B_\mu \Phi^\dagger - \frac{1}{ig_2}\partial_\mu \Phi \Phi^\dagger.
\label{newconnection}
\end{align}
This connection is ${\rm SU}(2)$-invariant. It however transforms under the ${\rm U}(1)$ transformations 
\begin{align}
W_\mu ~\mapsto~ e^{i \xi T_3} W_\mu e^{-i\xi T_3}- \frac{1}{g_2} \left(\partial_\mu\xi\right) T_3 \label{u1}
\end{align}

We can recognise in the ${\rm SU}(2)$-invariant connection $W_\mu$ a part of the quantity $D_\mu \Phi$. Indeed, we have
\be
\im (D_\mu \Phi )\Phi^\dagger = g_2 W_\mu - g_1 Y_\mu T^3.
\ee
Therefore we have
\begin{align}
{\rm Tr}|D_\mu \Phi|^2 = \frac{1}{2}\left( g_2^2 \left( W_{\mu}^1 W^{1\,\mu} + W_{\mu}^2 W^{2\,\mu} \right) + \left( g_2 W_\mu^3 -g_1 Y_\mu \right)^2\right)\label{DH2},
\end{align}
where we decomposed $W_\mu = W_\mu^a T^a$. These are the mass terms for the $W,Z$ bosons obtained from the kinetic term for the Higgs. Usually one obtains them by choosing a VEV for the Higgs and breaking the symmetry. Here we instead defined ${\rm SU}(2)$ invariant connection $W_\mu$, which appears in the covariant derivative acting on the ${\rm SU}(2)$ invariant doublets. The kinetic term for the Higgs then gives the mass terms for $W_\mu$ without any symmetry breaking. 

We can now define the usual linear combinations
\begin{align}
W^{\pm}_\mu \defeq \frac{1}{\sqrt{2}}\left( W_\mu^1 \mp i W_\mu^2\right),\qquad Z_\mu \defeq \frac{g_2W_\mu^3 - g_1 Y_\mu}{\sqrt{g_1^2+g_2^2}},
\end{align}
where the normalisation is a convention chosen for later convenience. By construction these fields are invariant under $SU(2)$ and transform under $U(1)$ as
\begin{align}
W_\mu^\pm ~\mapsto~ e^{\pm i\xi(x)}W_\mu^\pm, \qquad Z_\mu ~\mapsto~ Z_\mu
\end{align}
These three gauge fields are identified with the physical $SU(2)$ bosons which one can measure in an experiment. We can further define the Weinberg angle $\theta_W$ so that
\begin{align}
Z_\mu \defeq \cos(\theta_W)W_\mu^3 - \sin(\theta_W)Y_\mu
\end{align}
From this equation one can deduce (we will further motivate this choice later on) that the photon gauge field will be given by:
\begin{align}
A_\mu \defeq \sin(\theta_W)W_\mu^3 + \cos(\theta_W)Y_\mu
\end{align}
Notice that the field redefinition from $W^3_\mu, Y_\mu$ to $Z^\mu, A_\mu$ is an $SO(2)$ transformation.
\medskip
All in all, the Higgs sector Lagrangian can be rewritten in terms of physical quantities as follows:

\begin{align}\begin{split}
\mathcal{L}_{Higgs} &= -|D_\mu \phi|^2 - V\left( |\phi|^2 \right) \\
&=  -\left( \partial_\mu \rho \right)^2- \frac{(g_2\rho)^2}{2} \left( W^+ W^- +\frac{1}{2\cos^2(\theta_W)}Z_\mu Z^\mu\right) - V(\rho^2)\label{HiggsInter}
\end{split}\end{align} 

It is worth emphasising once again that to extract the mass terms for the $W,Z$ bosons no symmetry breaking was needed. The Higgs sector was merely reformulated in terms of the physical ${\rm SU}(2)$-invariant degrees of freedom of the theory. More details on this reparametrization of the Higgs field can be found in \cite{Faddeev:2008qc}.

\subsection{Yang-Mills sector}

We now perform the same change of variables in the Yang-Mills sector. One starts with the following Lagrangian:
\begin{align}
\mathcal{L}_{YM} = -\frac{1}{8}{\rm Tr}\left(B_{\mu\nu}B^{\mu\nu} \right)- \frac{1}{4}Y_{\mu\nu}Y^{\mu\nu},
\end{align}
where the curvature tensors are defined according to (\ref{faraday}). Since the field redefinition (\ref{newconnection}) is a gauge transformation, we can immediately write
\begin{align}
\mathcal{L}_{YM} = -\frac{1}{8}{\rm Tr}\left(W_{\mu\nu}W^{\mu\nu} \right)- \frac{1}{4}Y_{\mu\nu}Y^{\mu\nu}
\end{align}

It is now convenient to define the following curvature combinations:
\begin{align}
W_{\mu\nu}^\pm \defeq \frac{1}{\sqrt{2}}\left( W_{\mu\nu}^1 \mp i W_{\mu\nu}^2\right) \equiv D_\mu W^\pm_\nu - D_\nu W^\pm_\mu,
\end{align}
where the covariant derivatives are
\begin{align}
D_\mu W^\pm_\nu \equiv \left(\partial_\mu \pm ig_2 W_\mu^3\right)W_\nu^\pm.
\end{align}
We then have :
\begin{align}
{\rm Tr}\left(W_{\mu\nu}W^{\mu\nu} \right) = 2W^3_{\mu\nu}W^{3\,\mu\nu}+ 4 W^+_{\mu\nu}W^{-\,\mu\nu}.
\end{align}

Recall now that the $W^3, Y$ connections can be expressed in terms of the physical $Z, A$ connections as
\begin{align}
\left(\begin{array}{c}  W^3 \\ Y \end{array} \right) = \left(\begin{array}{cc}  \cos(\theta_W) & \sin(\theta_W) \\ -\sin(\theta_W) & \cos(\theta_W) \end{array} \right)\left(\begin{array}{c}  Z \\ A \end{array} \right)
\end{align}
Hence,
\begin{align}
D_\mu W^\pm_\nu \equiv \left(\partial_\mu \pm ie A_\mu \pm i g_2 \cos(\theta_W)Z_\mu\right)W_\nu^\pm,
\end{align}
where the electric charge $e$ is given by
\begin{align}
e \defeq g_2 \sin(\theta_W) = g_1 \cos(\theta_W).
\end{align}

Another expression that we need is
\be
W^3_{\mu\nu} = \cos(\theta_W) Z_{\mu\nu} +\sin(\theta_W)F_{\mu\nu} + \im g_2 \left(W_\mu^+ W_\nu^- - W^-_\mu W^+_\nu\right),
\ee
where $F_{\mu\nu}$ and $Z_{\mu\nu}$ are the curvature tensors of the photon and $Z$ boson. This gives the following final expression for the Yang-Mills sector Lagrangian:
\begin{align}\begin{split}
\mathcal{L}_{YM} = &-\frac{1}{4}F_{\mu\nu}F^{\mu\nu}-\frac{1}{4}Z_{\mu\nu}Z^{\mu\nu}-\frac{1}{2}W^+_{\mu\nu}W^{-\,\mu\nu} \\ & +\frac{g_2^2}{2}W_\mu^+W_\nu^-\left( W^{+\,\mu} W^{-\,\nu }-W^{-\,\mu} W^{+\,\nu} \right) \\& -ie \left( F_{\mu\nu} +\cot(\theta_W) Z_{\mu\nu} \right)W^{+\,\mu }W^{-\,\nu}.
\end{split}\end{align}

\section{Interactions}
\label{sec:inter}

In the previous section we have discussed what the Higgs and Yang-Mills sector Lagrangians become when written in terms of the ${\rm SU}(2)$ invariant variables. We have seen that the components $W^\pm_\mu, Z_\mu$ of the ${\rm SU}(2)$ invariant connection $W_\mu$ are massive fields, with the mass determined by the VEV of the scalar field $\rho$. We have also expressed the gauge sector Lagrangian in terms of the physical fields $W^\pm_\mu, Z_\mu, A_\mu$. We can now put all this to use and discuss how physical ${\rm SU}(2)$ invariant gauge fields interact with the physical ${\rm SU}(2)$ invariant fermions. The interaction vertices are different in the second order formulation. We only consider the quark sector, as leptons are analogous. 

\subsection{Weak interactions}

We first consider Higgsless interactions arising when the Higgs is taken to sit on its VEV $\rho=\nu$. We take the Lagrangian in the form (\ref{L2-quarks*}), where the Higgs field is absorbed into the fermionic fields. Since we assume here that Higgs is a constant, the covariant derivative $\D$ contains just the weak and electromagnetic connections. We have:
\begin{align}
\D_\mu Q_i = \left(\partial_\mu + \im g_2 W_\mu + \frac{\im g_1}{6} Y_\mu \right)Q_i,\quad \D_\mu \bar Q_i = \left(\partial_\mu + \im g_1\mathcal{Q}Y_\mu   \right)\bar Q_i,
\end{align}
where $\mathcal{Q}$ is the matrix of electric charges, which is in this case
\begin{align}
\mathcal{Q}\bar Q_i \equiv \left( \begin{array}{cc} -2/3&0\\ 0& 1/3\end{array}\right) \left( \begin{array}{c}\bar u_i\\\bar d_i \end{array}\right) .
\end{align}

We now rewrite everything in terms of the physical gauge-invariant fields. We have:
\begin{align}
 \im g_2W_\mu+ \frac{\im g_1}{6} Y_\mu = \frac{\im g_2}{\sqrt{2}} \left( \begin{array}{cc} 0&W^+_\mu\\ W^-_\mu& 0\end{array}\right)
+ \im e \mathcal{Q}A_\mu + \frac{\im e}{s_Wc_W}Z_\mu\left( \frac{1}{2}T^3-s_W^2\mathcal{Q}\right),\end{align}
where, as before, $ e \defeq g_1 \cos(\theta_W)$ and $s_W \equiv \sin(\theta_W),~c_W\equiv \cos(\theta_W)$. For the barred quarks we have:
\begin{align}
ig_1\mathcal{Q}Y_\mu = \im e\mathcal{Q}A_\mu - \im e\mathcal{Q} t_WZ_\mu ,
\end{align}
where $t_W\equiv \tan(\theta_W)$. Note that the electric charges of the unbarred quarks are opposite of those of the barred ones. Thus, the quark fields interact with the electromagnetic field in the usual way. 

Let us now consider the interactions with the $W$-bosons. The relevant part of the Lagrangian (\ref{L2-quarks*}) becomes
\be\label{W-quarks}
-\im \sqrt{2} g_2 \left( K^{ij} (\partial^A{}_{A'} \bar{u}_{i A}) W^{+ A'B} d_{j B} + (K^\dagger)^{ij} (\partial^A{}_{A'} \bar{d}_{i A}) W^{-A'B} u_{j B}\right),
\ee
where we reinstated the suppressed before spinor indices for clarity. The unitary matrix $K^{ij}$ is the Cabibbo-Kobayashi-Maskawa matrix that enters when the mass terms are diagonalised. Terms in (\ref{W-quarks}) give the interactions responsible for the $\beta$-decay, as well as for the mixing between the generations, thus making the more massive generations unstable. What is interesting is that in the second order formalism there is a derivative present in the interaction vertex. Of course, this can be seen to be the standard vertex with no derivative present if one uses the "reality condition" (\ref{quarks-RC}) to express the derivative of the barred spinors in terms of the hermitian conjugates of the unbarred. However, there is no need to introduce the primed spinors, and one can work with the Feynman rules that follow directly from (\ref{W-quarks}). 

The second order formalism also introduces new vertices quadratic in the gauge field. Indeed, we see that such vertices are present for both $A$ and $Z$ fields (but not for $W^\pm$). However, because the second order Lagrangian was obtained by a procedure of integrating out the primed 2-component spinors one is certain that the end results for the correlation functions of the unprimed fields are correctly reproduced. 

\subsection{Interactions with the Higgs}

A much more interesting story is with the Higgs field interactions, which are rather unusual in the second order formalism. Thus, we now consider fluctuations around the Higgs VEV $\rho=\nu +h(x)$.

Polynomial interactions with the bosons are the usual ones and can be read off from (\ref{HiggsInter}). The self-interactions of the Higgs are also as usual. Here we are specifically concerned with the fermionic sector that exhibits non-polynomiality in the Higgs. Let us again consider only the quark sector; for leptons everything is analogous. 

We recall that in the form of the Lagrangian (\ref{L2-quarks*}) the covariant derivative was defined so that:
\begin{align}
\frac{1}{\sqrt{\rho}}D ~\leftrightarrow~\mathcal{D}\frac{1}{\sqrt{\rho}}.
\end{align}
Therefore:
\begin{align}\label{new-D}
\mathcal{D} = D +\frac{1}{2}\partial \ln \rho.
\end{align}
This logarithmic non-polynomiality suggests that we should parametrize the Higgs field in a different way: 
\begin{align}
\rho \equiv \nu e^{\phi(x)}.
\end{align}
This shifts the non-polynomiality from the covariant derivative to the mass terms, schematically
\begin{align}
\rho^2 (\Lambda \bar Q) Q ~\rightarrow~ m^2 e^{2\phi(x)} \bar Q Q,
\end{align}
where $m$ is the quark mass. At the same time, the covariant derivative is now simple:
\begin{align}
\mathcal{D}Q \equiv \left(D + \frac{1}{2}\partial \phi(x)\right) Q.
\end{align}

The exponential non-polynomiality also enters into the Higgs with gauge fields interaction vertices, as well as in the kinetic term for the Higgs that now becomes:
\begin{align}
(\partial_\mu \rho)^2 = \nu^2 (\partial_\mu \phi)^2 e^{2\phi}.
\end{align}
We do not know which of these equivalent non-polynomial forms is better suited for practical calculations. At the same time, in practice one is only interested in terms involving just a few external Higgs lines. For calculations of this type it should matter little which parameterisation of the Higgs field is used. We will further return to the non-polynomiality issues in the last section.

\section{Unification}
\label{sec:unific}

In this section we discuss the unification in the second order formalism. 

\subsection{${\rm SU}(2)\times {\rm SU}(4)$ unification}

We first consider a Pati-Salam-like unification. Consider the second order Lagrangian in the form (\ref{L-intr}). Here the quarks' kinetic term contains a sum over the 3 colour indices. Spelling this out we have the following kinetic term:
\be
-\frac{2}{\rho} (D\bar{Q}^i)^r (DQ_i)^r - \frac{2}{\rho} (D\bar{Q}^i)^g (DQ_i)^g -\frac{2}{\rho} (D\bar{Q}^i)^b (DQ_i)^b -\frac{2}{\rho} D\bar{L}^i DL_i,
\ee
where $r,g,b$ are the 3 colours and $(DQ_i)^*$ denotes the projection of a triplet on a particular colour index. This form of the kinetic term cries for the interpretation of leptons as the fourth colour of quarks. This suggests that we put all of the SM fermions into two multiplets:
\be\label{uni-fermions}
S_i :=\left( \begin{array}{cccc} u_i^r & u_i^g & u_i^b & \nu_i \\ d_i^r & d_i^g & d_i^b & e_i \end{array}\right), \qquad 
\bar{S}_i :=\left( \begin{array}{cc} \bar{u}_i^r & \bar{d}_i^r\\ \bar{u}_i^g & \bar{d}_i^g \\ \bar{u}_i^b & \bar{d}_i^b \\ \bar{\nu}_i & \bar{e}_i \end{array}\right).
\ee
We can then rewrite the Lagrangian in terms of $S_i,\bar{S}_i$, using appropriate covariant derivatives for each field. However, what seems to spoil this picture is the different electric charges of the quarks and leptons. The same problem arises in the usual Pati-Salam treatment, where it is solved by using a non-trivial embedding of the SM gauge group into a larger group, see e.g. \cite{Baez:2009dj} for a nice exposition. 

Thus, to understand what is happening with the electric charges, we need to understand how the SM symmetry group sits inside some larger gauge group. Note that the symmetry group that is unbroken in the Lagrangian (\ref{L-intr}) is ${\rm U}(1)\times{\rm SU}(3)$. The weak ${\rm SU}(2)$ no longer acts on our fermions, as they are all ${\rm SU}(2)$-invariant objects. However, there is a leftover from this gauge group in the form of the massive gauge field (also ${\rm SU}(2)$-invariant) that acts on doublets $Q_i, L_i$ and does not act on $\bar{Q}_i, \bar{L}_i$. So, the group ${\rm SU}(2)$ is broken, but the fact that the fermions come as doublets tells us that it was there. Similarly, now that we put in (\ref{uni-fermions}) leptons on the same footing as the quarks, it appears that there is an ${\rm SU}(4)$ behind this construction. So, we take ${\rm SU}(2)\times {\rm SU}(4)$ as the GUT gauge group that acts on multiplets $S_i,\bar{S}_i$, and look for an embedding of ${\rm U}(1)\times{\rm SU}(3)$ into it. The sought embedding is given by
\be
{\rm U}(1)\times {\rm SU}(3) \ni \{\alpha, h\} \to \left\{ \left( \begin{array}{cc} \alpha^3 & 0\\ 0 & \alpha^{-3}\end{array} \right) , \left( \begin{array}{cc} \alpha h & 0 \\ 0 & \alpha^{-3} \end{array} \right)\right\}\in {\rm SU}(2)\times {\rm SU}(4).
\ee
Let us check how this works out for the charges. According to this prescription the ${\rm U}(1)$ acts on the up quarks as $u_i\to \alpha^{1+3} u_i$, which corresponds to the correct electric charge of $4/6=2/3$. Similarly, for the down quarks we have $d_i\to \alpha^{1-3} d_i$, which gives the correct electric charge of $-2/6=-1/3$. For the neutrino we have $\nu_i\to \alpha^{3-3} \nu_i$, which gives zero electric charge, and for the electrons $e_i\to \alpha^{-3-3} e_i$, which gives the electric charge $-1$. This gives all the correct quantum numbers of the unbarred fermions. For the barred ones it is clear that we simply have to use the hermitian conjugate representation of ${\rm SU}(2)\times {\rm SU}(4)$ (but not of the Lorentz group, because the barred fermions are still unprimed 2-component spinors). 

We can now write the kinetic terms for all the fermions in a very compact form
\be\label{unif-kin}
-\frac{2}{\rho} D\bar{S}^i DS_i.
\ee
Here $D$ is the relevant covariant derivative for each multiplet. The ${\rm SU}(3)$ and ${\rm U}(1)$ connections are present in both $DS_i$ and $D\bar{S}^i$ in a symmetric way, with the hermitian conjugate connections appearing in $D\bar{S}^i$. However, the massive ${\rm SU}(2)$ gauge field appears asymmetrically in that $D\bar{S}^i$ is diagonal in the isospin indices, while $DS_i$ is not. As we have already seen in the previous section, only $DS_i$ contains the $W^\pm$ gauge fields, while both $D\bar{S}^i$ and $DS_i$ contain the $Z$ field, albeit in an asymmetric way. This different coupling of the massive gauge fields to $\bar{S}_i, S_i$ is what causes the theory to be left-right asymmetric. 

Let us now discuss the mass terms. These can again be written in terms of $\bar{S}_i, S_i$ as
\be\label{unif-mass}
-\rho \, \bar{S}_i \Lambda^{ij} S_j.
\ee
The mass matrices $\Lambda^{ij}$ appearing here are complicated objects. Each of them is an $8\times 8$ block matrix that consists of 4 different entries $\Lambda^{ij}_u, \Lambda^{ij}_d, \Lambda^{ij}_\nu,\Lambda^{ij}_e$. It thus breaks ${\rm SU}(2)$ symmetry completely, while the ${\rm SU}(4)$ is broken down to ${\rm U}(1)\times {\rm SU}(3)$. 

Overall, the sum of two terms (\ref{unif-kin}), (\ref{unif-mass}) gives the Lagrangian (\ref{L-intr-unif}) quoted in the Introduction. We note that the unification described here is different from the Pati-Salam model, as no second ${\rm SU}(2)$ has been used. This seems natural in the second order formalism in which the weak ${\rm SU}(2)$ has been frozen from the beginning by using the gauge invariant variables. There is then no sense in introducing an independent copy of ${\rm SU}(2)$ that would act on the right-handed doublets, as in our framework these doublets simply transform under the hermitian conjugate representation of ${\rm SU}(2)\times {\rm SU}(4)$. It is also worth noting that ${\rm SU}(2)\times {\rm SU}(4)\sim {\rm SO}(3)\times {\rm SO}(6)\subset {\rm SO}(9)$. 

\subsection{${\rm SO}(10)$ unification}

It is known that {\it all} left-handed fermions of the SM can be put into a single 16-dimensional irreducible representation of ${\rm SO}(10)$, see e.g. \cite{Srednicki:2007qs},  section 97. So, our unprimed fermions $\bar{S}^i, S^i$ can be combined into a single multiplet of ${\rm SO}(10)$. This ${\rm SO}(10)$ is obtained from the already encountered ${\rm SU}(2)\times {\rm SU}(4)$ by adding another ${\rm SU}(2)$ that mixes $\bar{S}^i$ and $S^i$. Putting these groups together we have ${\rm SU}(2)\times{\rm SU}(2)\times{\rm SU}(4)\sim {\rm SO}(4)\times {\rm SO}(6)\subset {\rm SO}(10)$. The Lagrangian one obtains is of the same schematic form (\ref{L-intr-unif}), now with a single 16-dimensional fermionic multiplet ${\mathcal F}^i$. We will not explore this theory any further, as there seems to be no benefit in putting all fermions into a single multiplet from the point of view of reducing the complexity of the Lagrangian. At the same time, one has to introduce ${\rm dim}({\rm SO}(10))-{\rm dim}({\rm SU}(2)\times{\rm SU}(4))=45-18=27$ more gauge fields in the ${\rm SO}(10)$ unification as compared to the ${\rm SU}(2)\times {\rm SU}(4)$ framework. 

We also note that it is the ${\rm SO}(10)$ unification scheme that incorporates the ${\rm SU}(5)\subset{\rm SO}(10)$ viewpoint \cite{Georgi:1974sy} on the SM fermions. The second order version of the ${\rm SU}(5)$ model is also possible, but we refrain from spelling out the details, given that this model requires ${\rm dim}({\rm SU}(5))-{\rm dim}({\rm SU}(2)\times{\rm SU}(4))=6$ more gauge fields than ${\rm SU}(2)\times {\rm SU}(4)$ based GUT. It is thus less economical than the model described in the previous subsection. 

\section{Discussion}

In this paper we have rewritten the SM Lagrangian in a chiral form. The new Lagrangian involves only unprimed 2-component spinors and was obtained from the usual first order Lagrangian by integrating out the primed spinors. It is second order in derivatives. In the process, we were led to combine the ${\rm SU}(2)$ singlets into doublets, as well as to define ${\rm SU}(2)$-invariant combinations from the fermion doublets and the Higgs. Similarly, the angular part of the Higgs field was absorbed into the gauge fields to produce ${\rm SU}(2)$-invariant massive gauge fields. The bosonic part of this story is not new and has been explored by several authors. It appears that such a treatment of fermions is new. 

The obtained Lagrangian is quite compact, see (\ref{L-intr}) and can be used as the starting point for concrete computations of SM scattering amplitudes. It may be worth converting the arising computation rules into a computer code, as it is likely to lead to faster performance than the one based on the first order Lagrangian. The main reason for this expectation is that in the second order formulation the algebra of $\gamma$-matrices that needs to be performed is much simpler, as the propagator is a multiple of the identity matrix in the field space. 

Probably the most interesting aspect of the new Lagrangian (\ref{L-intr}) is that the Higgs field $\rho^2=|\phi|^2$ enters it non-polynomially. The analogy between the Higgs field ($\rho$ in our notation) and the conformal factor of the metric was already emphasised in \cite{Faddeev:2008qc}. Our rewrite of the SM Lagrangian strengthens this interpretation. Indeed, consider a conformal transformation $g_{\mu\nu} \to \rho^2 g_{\mu\nu}$. Under such a conformal rescaling the spinor metric $\epsilon_{AB}$ transforms as $\epsilon_{AB}\to\rho \epsilon_{AB}, \epsilon^{AB}\to\rho^{-1}\epsilon^{AB}$. The Dirac operator changes as $\nabla_{AA'} \chi_B \to \nabla_{AA'}\chi_B - (\partial_{BA'} \log\rho) \chi_A$, see e.g. \cite{Penrose:1987uia}, formula (5.6.15). Then, if we define the transformation rule for the spinors to be $S_A \to \rho^{-1} S_A$, similarly for $\bar{S}_A$, the quantity $\epsilon^{AB} D_{AA'} S_B$ transforms homogeneously $\epsilon^{AB} D_{AA'} S_B\to \rho^{-2} \epsilon^{AB} D_{AA'} S_B$, with the covariant derivative remaining unchanged. This implies that under such a transformation
\be
\sqrt{g} \,D\bar{S} DS \to \frac{1}{\rho} \sqrt{g} \, D\bar{S} DS,
\ee
where we have taken into account that $\sqrt{g}\to \rho^4 \sqrt{g}$, and there is an extra factor of $1/\rho$ coming from the contraction of the primed spinor indices. Similarly, 
\be
\sqrt{g}\, \bar{S} S \to \rho \sqrt{g}\, \bar{S} S.
\ee
Thus, we see that, as already observed in \cite{Faddeev:2008qc} for the bosonic sector, the Higgs field $\rho$ enters the fermionic Lagrangian (\ref{L-intr}) as the conformal factor of a transformation $g_{\mu\nu}\to\rho^2 g_{\mu\nu}$. 

Another potentially interesting outcome of our work is a relatively simple GUT theory based on ${\rm SU}(2)\times {\rm SU}(4)$. It suggests to group all (unbarred in our conventions) 2-component spinors into a single representation of this group, with the barred spinors combining into the complex conjugate representation. Given that the dimension of this gauge group is just $18$, the associated GUT is more economical than the better-known ${\rm SU}(5)$ and ${\rm SO}(10)$ theories. It would be interesting to explore this GUT model in more details. 

\section*{Acknowledgements} The authors were supported by an ERC Starting Grant 277570-DIGT. The second author also acknowledges support from the Alexander von Humboldt foundation, Germany, and from Max Planck Institute for Gravitational Physics, Golm.

\appendix 
\section{Curvature and covariant derivative conventions}

In this paper we dealing exclusively with unitary groups so that the inverse of a group object is its hermitian conjugate. Let $A_\mu \defeq A_\mu^s T^s$ be a connection gauge field. $T^s$ ($s=1,\ldots, {\rm{dim}(G)}$) are the generators of the Lie group, which we take to be hermitian, that satisfy:
\begin{align}
\left[ T^s, T^r \right] =i f^{srt}T^t
\end{align}
A vector $\phi$ in the fundamental representation transforms as:
\begin{align}
\phi~\mapsto~\phi^\Omega \equiv \Omega \phi,
\end{align}
where
\begin{align}
\Omega \equiv \exp \left(i  q\xi^s(x)T^s \right),\quad \xi^s(x)\in \mathbb{R}
\end{align}
and $q$ stands for the charge of the field while $\xi^s(x)$ are coordinates that parametrize the transformation.

The covariant derivative is constructed as follows
\begin{align}
D_\mu \phi \defeq \left(\partial_\mu +ig q A_\mu \right)\phi .
\end{align}
If we require that this transforms covariantly under the gauge transformations
\begin{align}
D_\mu \phi ~\mapsto ~\Omega D_\mu \phi
\end{align}
we deduce the transformation rule for the connection:
\begin{align}
A_\mu  ~\mapsto~ A_\mu^\Omega \equiv \Omega^\dagger A_\mu \Omega + \frac{1}{igq}\Omega^\dagger \left(\partial_\mu \Omega\right) ,\quad \Omega \in G
\end{align}
where $g$ denotes the coupling constant of the group. The field strength tensor or Yang-Mills curvature tensor is defined as:
\begin{align}
F_{\mu\nu} \defeq \partial_\mu A_\nu - \partial_\nu A_\mu + igq \left[A_\mu, A_\nu  \right]\label{faraday}
\end{align}
It transforms in the adjoint representation of the Lie group:
\begin{align}
F_{\mu\nu} ~\mapsto~ F_{\mu\nu}^\Omega \equiv \Omega^\dagger F_{\mu\nu} \Omega.
\end{align}


\begin{thebibliography}{99}

\bibitem{Feynman:1958ty} 
  R.~P.~Feynman and M.~Gell-Mann,
  ``Theory of Fermi interaction,''
  Phys.\ Rev.\  {\bf 109}, 193 (1958).
  
\bibitem{Brown:1958zz} 
  L.~M.~Brown,
  ``Two-Component Fermion Theory,''
  Phys.\ Rev.\  {\bf 111}, 957 (1958).

\bibitem{Cortes:1992wr}
  J.~L.~Cortes, J.~Gamboa and L.~Velazquez,
  ``Second order formalism for fermions,''
  Phys.\ Lett.\ B {\bf 313} (1993) 108
  [hep-th/9301071].
  
\bibitem{Morgan:1995te}
  A.~G.~Morgan,
  ``Second order fermions in gauge theories,''
  Phys.\ Lett.\ B {\bf 351} (1995) 249
  [hep-ph/9502230].
  

\bibitem{Chalmers:1997ui} 
  G.~Chalmers and W.~Siegel,
  ``Simplifying algebra in Feynman graphs, Part I: Spinors,''
  Phys.\ Rev.\ D {\bf 59}, 045012 (1999)
  [hep-ph/9708251].
    
\bibitem{Veltman:1997am}
  M.~J.~G.~Veltman,
  ``Two component theory and electron magnetic moment,''
  Acta Phys.\ Polon.\ B {\bf 29} (1998) 783
  [hep-th/9712216].

  
\bibitem{DelgadoAcosta:2010nx}
  E.~G.~Delgado-Acosta, M.~Napsuciale and S.~Rodriguez,
  ``Second order formalism for spin 1/2 fermions and Compton scattering,''
  Phys.\ Rev.\ D {\bf 83} (2011) 073001
  [arXiv:1012.4130 [hep-ph]].
  
\bibitem{Angeles:2011zz}
  R.~Angeles and M.~Napsuciale,
  ``Quantization of second order fermions,''
  J.\ Phys.\ Conf.\ Ser.\  {\bf 287} (2011) 012041.
  
\bibitem{AngelesMartinez:2011nt}
  R.~Angeles-Martinez and M.~Napsuciale,
  ``Renormalization of the QED of second order spin 1/2 fermions,''
  Phys.\ Rev.\ D {\bf 85} (2012) 076004
  [arXiv:1112.1134 [hep-ph]].
    
\bibitem{VaqueraAraujo:2012qa}
  C.~A.~Vaquera-Araujo, M.~Napsuciale and R.~Angeles-Martinez,
  ``Renormalization of the QED of Self-Interacting Second Order Spin 1/2 Fermions,''
  JHEP {\bf 1301} (2013) 011
  [arXiv:1205.1557 [hep-ph]].
   
  
 

\bibitem{Vlasov:1987vt} 
  V.~V.~Vlasov, V.~A.~Matveev, A.~N.~Tavkhelidze, S.~Y.~.Khlebnikov and M.~E.~Shaposhnikov,
  ``Canonical Quantization Of Gauge Theories With Scalar Condensate And The Problem Of Spontaneous Symmetry Breaking. (in Russian),''
  Fiz.\ Elem.\ Chast.\ Atom.\ Yadra {\bf 18}, 5 (1987).
 
\bibitem{Chernodub:2008rz}
  M.~N.~Chernodub, L.~Faddeev and A.~J.~Niemi,
  ``Non-abelian Supercurrents and Electroweak Theory,''
  JHEP {\bf 0812} (2008) 014
  [arXiv:0804.1544 [hep-th]].
  
  
\bibitem{Faddeev:2008qc}
  L.~D.~Faddeev,
  ``An Alternative interpretation of the Weinberg-Salam model,''
  arXiv:0811.3311 [hep-th].

\bibitem{Masson:2010vx}
  T.~Masson and J.~-C.~Wallet,
  ``A Remark on the Spontaneous Symmetry Breaking Mechanism in the Standard Model,''
  arXiv:1001.1176 [hep-th].
  
\bibitem{Pati:1974yy} 
  J.~C.~Pati and A.~Salam,
  ``Lepton Number as the Fourth Color,''
  Phys.\ Rev.\ D {\bf 10}, 275 (1974)
  [Erratum-ibid.\ D {\bf 11}, 703 (1975)].
    
\bibitem{TorresGomez:2012sr}
  A.~Torres-Gomez and K.~Krasnov,
  ``Fermions via spinor-valued one-forms,''
  arXiv:1212.3452 [hep-th].


\bibitem{Dreiner:2008tw} 
  H.~K.~Dreiner, H.~E.~Haber and S.~P.~Martin,
  ``Two-component spinor techniques and Feynman rules for quantum field theory and supersymmetry,''
  Phys.\ Rept.\  {\bf 494}, 1 (2010)
  [arXiv:0812.1594 [hep-ph]].


\bibitem{Asaka:2005pn} 
  T.~Asaka and M.~Shaposhnikov,
  ``The nuMSM, dark matter and baryon asymmetry of the universe,''
  Phys.\ Lett.\ B {\bf 620}, 17 (2005)
  [hep-ph/0505013].

\bibitem{Baez:2009dj} 
  J.~C.~Baez and J.~Huerta,
  ``The Algebra of Grand Unified Theories,''
  Bull.\ Am.\ Math.\ Soc.\  {\bf 47}, 483 (2010)
  [arXiv:0904.1556 [hep-th]].
  
\bibitem{Srednicki:2007qs} 
  M.~Srednicki,
  ``Quantum field theory,''
  Cambridge, UK: Univ. Pr. (2007) 641 p
  
\bibitem{Penrose:1987uia} 
  R.~Penrose and W.~Rindler,
  ``Spinors and Space-Time: Volume 1 : Two-Spinor Calculus and Relativistic Fields,''

\bibitem{Georgi:1974sy} 
  H.~Georgi and S.~L.~Glashow,
  ``Unity of All Elementary Particle Forces,''
  Phys.\ Rev.\ Lett.\  {\bf 32}, 438 (1974).



\end{thebibliography}
\end{document}